\documentclass[showpacs,prb,superscriptaddress]{revtex4}
\usepackage{epsfig }
\usepackage{amsmath}
\usepackage{amssymb}

\newcommand{\p}{\partial}

\newcommand{\ph}{\varphi}
\newcommand{\m}{\mathcal}

\begin{document}

\title{Quasi-bound  states of Schr\"{o}dinger and Dirac electrons in magnetic quantum dot.}
\author{M.~Ramezani Masir}\email{mrmphys@gmail.com}
\affiliation{Departement Fysica, Universiteit Antwerpen \\
Groenenborgerlaan 171, B-2020 Antwerpen, Belgium}
\author{A. Matulis}\email{amatulis@takas.lt}
\affiliation{Departement Fysica, Universiteit Antwerpen \\
Groenenborgerlaan 171, B-2020 Antwerpen, Belgium}
\affiliation{Semiconductor Physics Institute, Go\v{s}tauto 11,
LT-01108 Vilnius, Lithuania}
\author{F.~M.~Peeters}\email{francois.peeters@ua.ac.be}
\affiliation{Departement Fysica, Universiteit Antwerpen \\
Groenenborgerlaan 171, B-2020 Antwerpen, Belgium}
\affiliation{Departamento de F\'{\i}sica, Universidade Federal do\\
Cear\'{a}, Caixa Postal 6030, Campus do Pici, 60455-760 Fortaleza,
Cear\'{a}, Brazil.}
\begin{abstract}
The properties of a two-dimensional electron are investigated in the
presence of a circular step magnetic field profile. Both electrons
with parabolic dispersion as well as Dirac electrons with linear
dispersion are studied. We found that in such a magnetic quantum dot
no electrons can be confined. Nevertheless close to the Landau
levels quasi-bound states can exist with a rather long life time.
\end{abstract}


\date{\today}

\maketitle

\twocolumngrid

\section{Introduction}
During the last five years, graphene (a single layer of carbon
atoms) has become a very active field of research in nanophysics
\cite{no05,zh05}. It is expected that this material will serve as a
base for new electronic and opto-electric devices. One of the most
challenging task is to learn how to control the electron behavior
using electric fields in this two-dimensional (2D) layer. This task
is made complicated by  the so called Klein effect according to
which Dirac electrons in graphene can tunnel through any electric
barrier \cite{ka06}. As a consequence in electrically created
quantum dots there are no bound states but only quasi-bound states,
or so called resonances \cite{ho07,ma08,he08} which, however, under
certain conditions can have a long life time.

An alternative approach to control the motion of electrons is to use
non homogeneous magnetic fields which can be created e.g. through
the deposit of nonostructured ferromagnets. \cite{pe93,Rei1,Rei2}.
Recently it was shown that nonhomogeneous magnetic structures are
able to confine Dirac electrons in graphene
\cite{rm08,egg,Park,Rakyta,Ghosh,zhai,MT,Xu}. However, up to now
semi-infinite (homogeneous in one direction) structures were
considered, what makes the analysis more simple because the problem
is reduced into a one-dimensional one.

In this paper we consider a finite size magnetic structure where the
magnetic field is nonzero only in a finite region of space. Namely,
we consider a model homogeneous magnetic field that is non-zero in a
circle that we call the magnetic dot. This situation is the inverse
of the one considered in Ref. 11 where a magnetic anti-dot was
considered, as in Ref. 9 for the case of  normal electrons, where
the magnetic field is zero in a circular region and non-zero outside
this region. Such a model system can be realized by having a
magnetic vortex piercing the graphene layer or by overlaying
graphene with type I superconductor with a circular hole placed in
the perpendicular magnetic field. In order to reveal the
peculiarities of the behavior of Dirac electrons in such magnetic
dot we compare the result with those for standard electrons with parabolic
dispersion law.

We show that it is impossible to confine 2D electrons in a magnetic
dot in contrast to semi-infinite magnetic structures neither in the
case of graphene nor in the case of the standard electron, and
consequently, all Landau levels convert themselves into unbound
states. Nevertheless, long living quasi-bound states can be present.
We studied them using the local density of states technique applied
previously for the investigation of electrically confined electrons
\cite{ma08}.

The paper is organized as follows. In section II the model of a
magnetic dot for a standard electron is considered. The problem is
formulated in subsection A, the local density of states technique is
presented in subsection B, the results are discussed in subsection
C, while in subsection D the complex energy eigenvalues of the
problem are described. In the corresponding subsections of section
III the problem of the Dirac electron in a magnetic dot is
presented. Our conclusions are given in section IV.

\section{Electron with parabolic energy dispersion}

\begin{figure}[ht]
\begin{center}
\includegraphics[width=5cm]{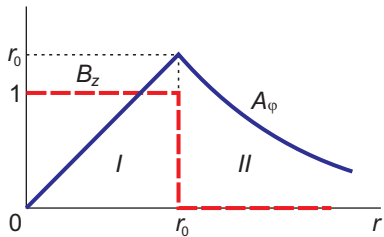}
\caption{(Color online) Azimuthal vector potential component
$A_{\ph}$ (blue solid curve), and perpendicular magnetic field $B_z$
(red dashed curve) as functions of the radial coordinate.}
\label{fig1}
\end{center}
\end{figure}

We assume that a homogeneous magnetic field ${\bf B}_0$ is present
in a circular area of radius $r_0$, while there is no magnetic field
outside it, namely, ${\bf B}_0({\bf r}) = {\bf e}_zB_0\Theta(r_0 -
r)$. The behavior of the electron is described by the stationary
Schr\"{o}dinger equation
\begin{equation}\label{sse}
  \{H - E\}\Psi({\bf r}) = 0,
\end{equation}
with the Hamiltonian
\begin{equation}\label{hamst}
  \displaystyle{H = - \frac{1}{2}\left(\nabla + i{\bf A}\right)^2.}
\end{equation}
Because of the cylindric symmetry of the problem we choose the
symmetric gauge for the vector potential defining its single
azimuthal component as
\begin{equation}\label{vp}
  A_{\ph}(r) \equiv = \frac{1}{2}\left\{
  \begin{array}{ll}r, & r < r_0; \\ r_0^2/r, & r_0 < r. \end{array}\right.
\end{equation}
This azimuthal component is shown in Fig.~\ref{fig1} together with
the magnetic field profile.
In order to simplify the notations we use dimensionless variables,
based on the magnetic field strength value $B_0$. Thus, the magnetic
field $B({\bf r})$ is measured in $B_0$ units, all distances are measured
in the unit of magnetic length $l_B = \sqrt{c\hbar/eB_0}$, energy
and potential in $\hbar\omega_c$ ($\omega_c = eB_0/mc$), and
vector potential in $B_0l_B$ units. In the case of electron moving
at a GaAs/AlGaAs interface ($m^*=0.067$) and magnetic field of 1\ T
the unit of length is $l_B=250$ nm, and the energy unit is $20$ meV.

\subsection{Solution of eigenvalue problem}

The Schr\"{o}dinger equation (\ref{sse}) in cylindric coordinates
reads
\begin{equation}\label{srccyl}
  \left\{ \frac{1}{r}\frac{\p}{\p r}r\frac{\p}{\p r}
  + \frac{1}{r^2}\frac{\p^2}{\p\ph^2} + \frac{iA_{\ph}}{r}\frac{\p}{\p\ph}
  - A_{\ph}^2 + 2E\right\}\Psi = 0.
\end{equation}
Substituting the wave function
\begin{equation}\label{wfrad}
  \Psi \equiv \Psi(r,\ph) = e^{im\ph}\psi(r)
\end{equation}
we arrive at the radial equations
\begin{subequations}\label{srccylrad}
\begin{eqnarray}
\label{srccylrad1}
  \left\{ \frac{1}{r}\frac{d}{dr}r\frac{d}{dr}
  - \left(\frac{m}{r} + \frac{r}{2}\right)^2  + 2E\right\}\psi_I(r) = 0, \\
\label{srccylrad2}
  \left\{ \frac{1}{r}\frac{d}{dr}r\frac{d}{dr}
  - \frac{(m + r_0^2/2)^2}{r^2} + 2E\right\}\psi_{II}(r) = 0,
\end{eqnarray}
\end{subequations}
which have to be solved inside the dot (region $I$) and outside it (region $II$).
The boundary conditions (the continuity of the wave function and its radial derivative) have
to be satisfied at the dot border ($r=r_0$).

The regular solution inside the dot can be expressed via the
confluent hypergeometric function (Kummer function $M(a|c|z)$):
\begin{equation}\label{solin}
\begin{split}
  \psi_I(r) &= Af(r) = Ar^{|m|}e^{-r^2/4} \\
&  \times M\left((|m|+m)/2+1/2-E\big||m|+1\big|r^2/2\right),
\end{split}
\end{equation}
while the solution outside it is composed of two Bessel functions
\begin{equation}\label{solout}
  \psi_{II}(r) = BJ_{\nu}(kr) + CY_{\nu}(kr),
\end{equation}
where $k=\sqrt{2E}$ is the momentum of the free electron (measured
in $l_B^{-1}$ units), and $\nu = m+r_0^2/2$. Note both functions
($J_{\nu}$ and $Y_{\nu}$) suit us, as they vanish in the limit
$r\to\infty$.

\begin{figure}[ht]
\begin{center}
\includegraphics[width=5.5cm]{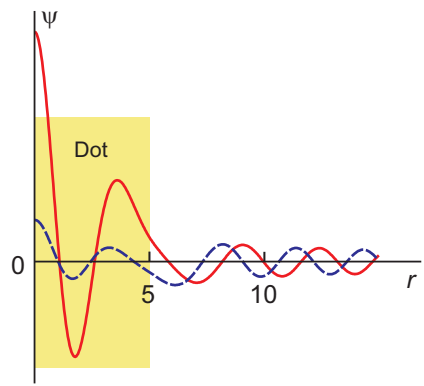}
\caption{(Color online) The wave functions for $m=0$, $r_0=5$: $E
= 2.5$ -- red solid curve, and $E = 2.8$ -- dashed blue curve. The
magnetic dot region is indicated by shadowed rectangle.}
\label{fig2}
\end{center}
\end{figure}

Thus, we have three constants $A$, $B$, and $C$. They can not be
defined from the above mentioned two boundary conditions. That is
why we have to conclude that there are no bound states, and
consequently, a magnetic field in a finite region of the 2D plane
can not confine the electron. However, quasi-bound states can be
expected when the electron energy in the dot is close to the Landau
levels with energy
\begin{equation}\label{ll}
  E_{n,m} = n + \frac{|m| + m + 1}{2}
\end{equation}
(here $n=0,1,\cdots$ and $m=0,\pm 1,\cdots$) defined in the case of
homogeneous magnetic field. Confirmation of this statement follows
from Fig.~\ref{fig2}, where the electron wave functions for two
different energies are shown.

We see that in the case of $E=2.5$ (red solid curve) which
corresponds to the Landau level with $n=2$ and $m=0$ the wave
function is large in the dot region (shown in Fig.~\ref{fig2} by
shadowed yellow rectangle), while for the case of energy $E=2.8$ ,
which does not coincide with any Landau level energy, it does not
have any appreciable large value inside the dot, and actually does
not differ much from the wave function for a free electron
calculated in cylindric coordinates.

\subsection{Local density of states}
\label{sec_lds}

Next we will look for possible long living quasi-stationary states
in the magnetic dot. In principle such quasi-bound (or
quasi-stationary) states have to be described by the solution of the
time dependent Schr\"{o}dinger equation which is much more
complicated as compared with the standard eigenvalue problem. There
are, however, several alternative approaches which enable us to
investigate properties of quasi-bound states by stationary means. We
follow the method presented in detail in Ref.~\onlinecite{ma08}, and calculate
the local density of states. The basic idea is to confine the
electron in a large region of finite radius $R$, where its wave
function obeys the zero boundary condition at the border ($r=R$)
and treat the problem as a stationary one. A measurement that probes
quantum dot properties, say, measuring of the tunnelling current
directed perpendicular to the dot with STM, or power absorption in
near field infrared spectroscopy, has to depend on the averaged
value of the electron wave function in the dot. Therefore we
introduce the integral
\begin{equation}\label{lds}
  \m{I}(E) = 2\pi\int_0^{\infty}rdr F(r)|\Psi(r)|^2,
\end{equation}
%\
which depends on the electron wave function, and actually is
proportional to the so called \textit{local density of states}. The
aperture function $F(r)$ characterizes the interaction of the
electron with the measuring probe.

This integral is sensitive to the probability to find the electron
in the dot, and in the case of a quasi-bound state it will exhibit a
peak corresponding to the energy of this state. The width of the
peak is related to the inverse of the life time of this
quasi-stationary state.

For the sake of determinacy we use the aperture function of a
gaussian:
\begin{equation}\label{gtf}
  F(r) = br_0^2 e^{-br^2}, \quad b = r_0^{-2}\ln 10.
\end{equation}
which corresponds to the probability to find the electron in the dot
area $\pi r_0^2$. In the case of larger $b$ value instead of the
local density of states we obtain the squared wave function value in
the center of the dot, while in the case of smaller $b$ value the
peculiarities of the dot are washed out.

The solution of the Schr\"{o}dinger equation (\ref{srccylrad}) given
by Eqs.~(\ref{solin}) and (\ref{solout}) has to satisfy the
following boundary conditions:
\begin{subequations}\label{bc}
\begin{eqnarray}
\label{bc1}
  \psi_I(r_0) &=& \psi_{II}(r_0), \\
\label{bc2}
  \psi_{I,r}(r_0) &=& \psi_{II,r}(r_0), \\
\label{bc3}
  \psi_{II}(R) &=& 0,
\end{eqnarray}
\end{subequations}
which converts our problem into an eigenvalue problem.
Here and further the subscript $_{r}$ means the derivative over $r$.

At the end, we are interested in the limiting case $R\to\infty$.
Therefore, in the last of Eqs.~(\ref{bc}) we replace the Bessel
functions by their asymptotic, namely, we have
\begin{equation}\label{bcas}
\begin{split}
&  B\cos(kR - \ph_m) + C\sin(kR - \ph_m) = 0, \\
&  \ph_m = \pi\left\{m + (r_0^2+1)/2\right\}/2.
\end{split}
\end{equation}
instead of Eq.~(\ref{bc3}).

Postponing till later the proper wave function normalization we
assume that $B = \cos\Phi$ and $C = \sin\Phi$ and rewrite the above
equation as
\begin{equation}\label{bc3f}
  \cos(kR - \ph_m - \Phi) = 0.
\end{equation}
This equation shows that the eigenvalues of the considered problem
are approximately separated by $\Delta k = \pi/R$, and reduce to a
continuum spectrum in the limit $R\to\infty$. Constructing some
averaged description which is valid when calculating the local
density of states, we replace Eq.~(\ref{bc3}) by the following
one:
\begin{equation}\label{norm}
  B^2 + C^2 = 1.
\end{equation}
Now solving it together with Eqs.~(\ref{bc1},b) we obtain three
constants:
\begin{equation}\label{constabc}
  A = - \frac{2}{\pi r_0}W, \quad B = WQ, \quad C = -WP,
\end{equation}
with
\begin{subequations}\label{constpq}
\begin{eqnarray}
  P &=& J_{\nu}f_r - J_{\nu,r}f, \\
  Q &=& Y_{\nu}f_r - Y_{\nu,r}f, \\
  W &=& \left(P^2 + Q^2\right)^{-1/2}.
\end{eqnarray}
\end{subequations}
The obtained constants enable us to calculate the
integral (\ref{lds}).

In order to convert the above integral into the local density of
states we have to multiply it by two additional constants. One of
them is the wave function normalization factor $N$, which can be
estimated calculating the integral of the squared wave function in
the limit of large radius $R$. The replacement of the Bessel
functions by their asymptotic immediately leads to $N = k / 2R$. The
second one is a consequence of the replacement of the summation over
the discrete eigenvalues by the integration over energy, which is
given by the factor $R/\pi k$. Together they give $1/2\pi$, which
results into definition of the local density of states
\begin{equation}\label{stlds}
  \rho(E) = \frac{1}{2\pi}\m{I}(E).
\end{equation}

\subsection{Numerical results}

We solved numerically Eqs.~(\ref{constpq}). Inserting the obtained
results into (\ref{constabc}), and later in Eqs.~(\ref{solin}) and
(\ref{solout}) we obtained the wave function what enabled us to
calculate the integral (\ref{lds}), and finally the local density of
states (\ref{stlds}).

\begin{figure}[ht]
\begin{center}
\includegraphics[width=7cm]{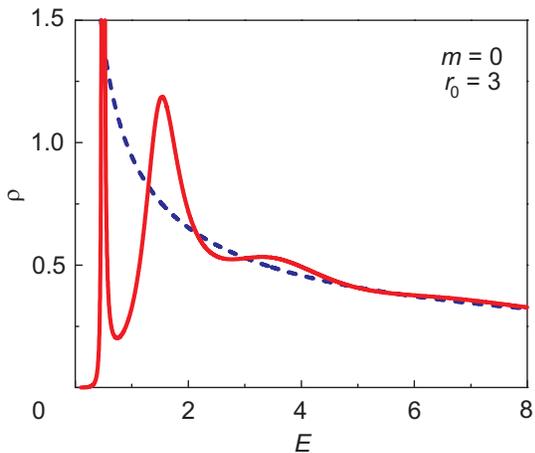}
\caption{(Color online) The local density of states for $m=0$ and
$r_0=3$ shown by the solid red curve. The same density calculated
for a free electron according to Eq.~(\ref{sfrlds}) is shown by the
blue dashed curve.} \label{fig3}
\end{center}
\end{figure}

A typical result for the local density of states as a function of
electron energy is shown in Fig.~\ref{fig3}.
We clearly see peaks close to the energies of the Landau levels
(\ref{ll}) calculated for the case of a homogeneous magnetic field.
These peaks are broadened indicating that they are not really bound
states in the magnetic dot. The broadening is larger for higher
energy peaks.

The next thing which also is seen in Fig.~\ref{fig3} is a decreasing
background with energy. This background is due to the states of the
free electron in the absence of the magnetic dot. To justify this
statement we made the same averaging (over circle of radius $r_0$)
with the same gaussian aperture function (\ref{gtf}) of the radial
component of the free electron wave function (when there is no
magnetic dot). This function reads
\begin{equation}\label{solfree}
  \psi_{\mathrm{free}}(r) = J_{m}(kr)
\end{equation}
and is valid in the whole 2D plane. Inserting this function into
integral (\ref{lds}) and later in Eq.~(\ref{stlds}) and using Tables
of integrals \cite{grad00} we obtain the local density of states for
a free electron
\begin{equation}\label{sfrlds}
  \rho_{\mathrm{free}}(E) = \frac{r_0^2}{2}e^{-E/b}I_m(E/b),
\end{equation}
where $I_m(x)$ stands for the modified Bessel function of the first
kind. This local density of free electron in the case of $m=0$ is
shown in the same Fig.~\ref{fig3} by the blue dashed curve.
Comparing these two curves we clearly see how increasing the
electron energy we reduce the influence of the magnetic dot on the
electron behavior, and the local density of states converts itself
gradually into the free electron one.

\begin{figure}[ht]
\begin{center}
\includegraphics[width=8cm]{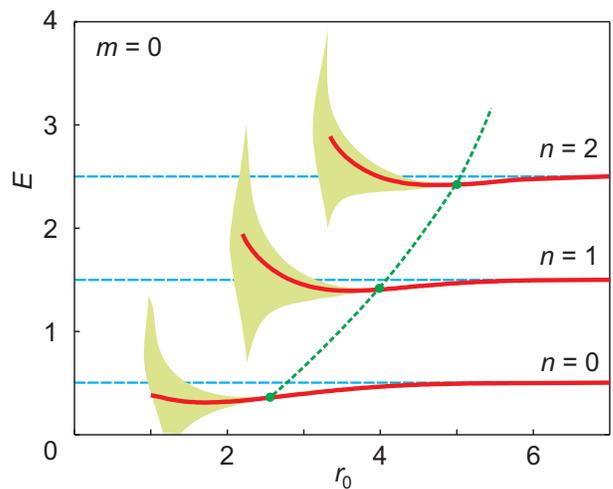}
\caption{(Color online) Quasi-bound states with orbital momentum
$m=0$. The energies of these states are given by red solid curves and
the widths (i.~e.~the inverse of the life time) by shadowed
regions. The Landau levels are indicated by blue dashed lines.}
\label{fig4}
\end{center}
\end{figure}

\begin{figure}[ht]
\begin{center}
\includegraphics[width=8cm]{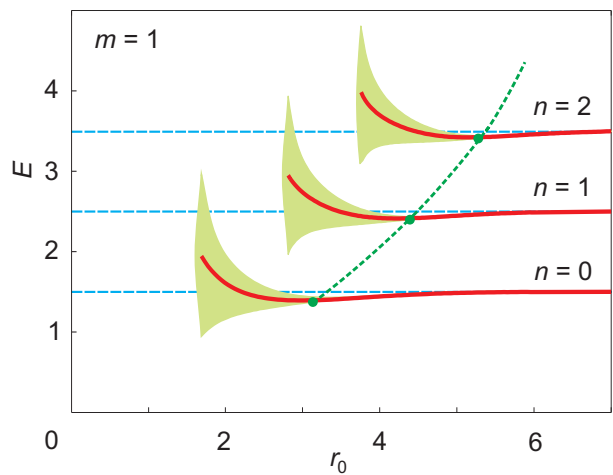}
\caption{(Color online) The same as Fig.~\ref{fig4} but now for
$m=1$.} \label{fig5}
\end{center}
\end{figure}

We fitted the peaks in the density of states by Lorentzian functions
$a_n\gamma_n/\{(E-E_n)^2 + \gamma_n^2\}$ defining three parameters
for any of them: the position $E_n$, its broadening $\gamma_n$, and
the amplitude $a_n$. Two of them (the position and broadening) are
shown in Figs.~\ref{fig4} and \ref{fig5} for different orbital
momenta as functions of the radius of the dot $r_0$. The position of
the quasi-bound states $E_n$ are shown by the red solid curves while
the broadening of the peaks is indicated by the shadowed areas
limited by the $E_n\pm\gamma_n$ curves.

Notice that the levels to the right of the green dotted curve are
extremely narrow and their position coincides with the Landau levels
(\ref{ll}) shown by the blue dashed horizontal lines. In fact this
means that almost all electron wave function is located in the
magnetic dot (using the classical description language we may say
that the electron rotates along the Larmor circle inside the dot)
and it does not touch the border of the magnetic dot. When the dot
radius $r_0$ becomes smaller the Larmor circle touches the dot
border and tunnelling of the electron outside the dot starts which
broadens the level. The partial penetration of the wave function
outside the dot leads to a lowering of the quasi-bound state energy
as well. The raising of this energy for small $r_0$ values is caused
by the large asymmetry of the peak where actually the approximate
replacement of the peak by a Lorentzian type function is no longer
valid. This picture is more or less the same for all positive $m$
values (compare Figs.~\ref{fig4} and \ref{fig5}). The difference is
that for larger $m$ values the levels start at higher energies what
is in agreement with the expression for Landau levels (\ref{ll}).

\begin{figure}[ht]
\begin{center}
\includegraphics[width=8cm]{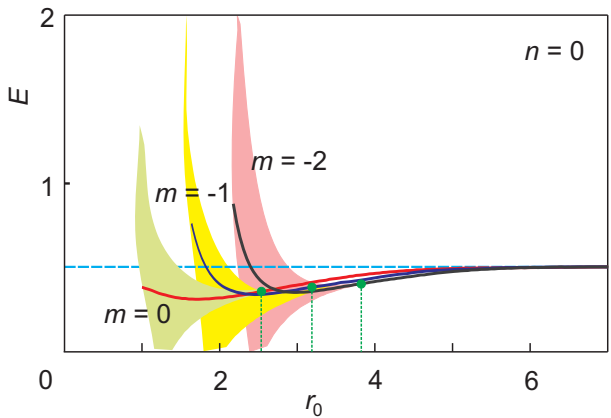}
\caption{(Color online) The lowest quasi-bound state with $n=0$ and
different negative $m$ values. The vertical dotted green lines are
the analog of the dotted curve in Figs.~\ref{fig4} and \ref{fig5},
separating the weakly broadened states from those with small
lifetime.} \label{fig6}
\end{center}
\end{figure}

The picture for negative $m$ values is different as shown in
Fig.~\ref{fig6}.
All of them belong to the same Landau level energy which is an
expression of the degeneracy of the Landau level. We see that with
decreasing radius of the dot $r_0$ the levels with different $m$
disappear step by step, the ones with smaller absolute $m$ values
disappear later. This is in agreement with the fact that the larger
the  $|m|$ value the larger the radius of the electron trajectory,
and the electron wave function is closer to the dot edge.

The increase of the peak broadening at small $r_0$ values is so
steep that it is worth to divide all the peaks into two classes as
 shown in the above figures by the green dotted curves. The
levels on the left side of these curves belong to essentially
broadened quasi-bound states, while those on the right side from the
experimental point of view can hardly be distinguished from the real
bound states.

One can rudely estimate the position of this dividing curve
comparing the approximate dimensions of the electron wave function
calculated in the case of a homogeneous magnetic field (which
actually coincides with function (\ref{solin})) with the magnetic
dot radius $r_0$. A more accurate estimation can be obtained solving
the stationary Schr\"{o}dinger equation for complex energy
eigenvalues as was described in Ref.~\onlinecite{he08}. We draw these dividing
curves in Figs.~\ref{fig4}-\ref{fig6} using this technique which is
sketched in the following section.

\subsection{Complex energy technique}
\label{compentex}

According to Ref.~\onlinecite{he08} the lifetime of the quasi-bound state,
or the trapping time of the electron in the quantum dot can be
estimated solving the time-independent Schr\"{o}dinger equation and
applying boundary conditions of the outgoing waves at the sharp dot
border.

In our case this technique reduces to connecting the wave function
(\ref{solin}) defined in the dot with the outgoing electron wave
function outside it, which is given by the first kind Hankel function
\begin{equation}\label{hwf}
  \psi_{\mathrm{out}}(r) = H_{\nu}^{(1)}(kr) = J_{\nu}(kr)
  + iY_{\nu}(kr)
\end{equation}
with
\begin{equation}\label{niudef}
  \nu = m+r_0^2/2.
\end{equation}
Applying the boundary conditions for these wave functions and their
derivatives we obtain the following equation:
\begin{equation}\label{ece}
  f_r(r_0)H(kr_0) - f(r)H_r(kr_0) = 0,
\end{equation}
where for the sake of simplicity we omitted the indexes of the
Hankel functions. The indexes which are only left indicates the
derivative over the coordinate $r$. This equation has to be solved
for a complex energy (or complex $k$), the imaginary part of the
energy gives the inverse of the lifetime.

For finding the dotted curves in Figs.~\ref{fig4}-\ref{fig6}
separating the quasi-bound and nearly bound states it is enough to
solve the above equation by means of a perturbation expansion in
terms of the momentum difference $\Delta k = k - k_0$ where $k_0 =
\sqrt{2E_{n,m}}$ with energy $E_{n,m}$ of the unperturbed Landau
level. Limiting ourselves to first order in $\Delta k$ we arrive at
the following expression:
\begin{equation}\label{momdev}
  \Delta k = \frac{H_rf - Hf_r}{Hf_{r,k} + H_kf_r - fH_{r,k} - H_rf_k}.
\end{equation}
All functions and their derivatives over $r$ and $k$ have to be
calculated at $r=r_0$ and $k=k_0$.

Now introducing the energy deviation from the Landau level energy
\begin{equation}\label{endev}
  \Delta E = E - E_{n,m} \approx k_0\Delta k,
\end{equation}
taking its imaginary part and equating it to $10^{-2}$ (it is
expected that a smaller broadening can hardly be revealed
experimentally) we obtained the points connected by the green dotted
curve in Figs.~\ref{fig4}-\ref{fig6} separating the quasi-bound
states from nearly bound states.

\section{Dirac electron in graphene}

Now we repeat the above calculation for the magnetic dot applying it
to the case of a Dirac electron in graphene, where the low-energy
quasi-particles (electrons and holes) are described by the following
dimensionless Dirac-like Hamiltonian:
\begin{equation}\label{dham}
  H = \boldsymbol{\sigma}\left(-i\nabla + {\bf A}\right).
\end{equation}
Here, $\boldsymbol{\sigma}=\{\sigma_x,\sigma_y\}$ stands for the
$2\times 2$ Pauli matrices. The units are based on the magnetic
field strength $B_0$ and they are the same as in previous section,
except the unit of energy which now is $v_F\hbar/l_B$ with the Fermi
velocity $v_F=10^8$ cm s$^{-1}$. In the case of a 1 T magnetic field
this energy unit is 2.6 meV. The vector potential is given by
Eq.~(\ref{vp}).

\subsection{Solution of eigenvalue problem}

The approach is based on the same stationary Schr\"{o}dinger
equation (\ref{sse}) but now with the matrix Hamiltonian
(\ref{dham}), which results into a set of two differential equations.
Assuming the wave function of the following form:
\begin{equation}\label{wfdir}
  \Psi = e^{im\ph}\begin{pmatrix} a(r) \\ ie^{i\ph}b(r) \end{pmatrix},
\end{equation}
we arrive at a set of two equations for the radial wave function
components
\begin{subequations}\label{twosysrad}
\begin{eqnarray}
\label{twosysrad1}
  \left\{\frac{d}{dr} + A(r) + \frac{m+1}{r}\right\}b &=& Ea, \\
\label{twosysrad2}
  -\left\{\frac{d}{dr} - A(r) - \frac{m}{r}\right\}a &=& Eb,
\end{eqnarray}
\end{subequations}
which has to be solved in the two regions ($I$ in the dot, and $II$
--- outside it). We require the continuity of the obtained
components at the dot border $r_0$
\begin{equation}\label{dbc}
  a_I(r_0) = a_{II}(r_0), \quad b_I(r_0) = b_{II}(r_0).
\end{equation}

Instead of solving these first order differential equations it is
more convenient to convert them into second order differential
equations for a single component, say for component $b$
\begin{subequations}\label{equa2}
\begin{eqnarray}
  \left\{\frac{1}{r}\frac{d}{dr}r\frac{d}{dr} - \frac{(m+1)^2}{r^2}
   - \frac{r^2}{4} + \left[E^2 - m\right]\right\}b_I = 0, \phantom{m}\\
  \left\{\frac{1}{r}\frac{d}{dr}r\frac{d}{dr} + \left[E^2
  - \frac{(m + 1 + r_0^2/2)^2}{r^2}\right]\right\}b_{II} = 0. \phantom{m}
\end{eqnarray}
\end{subequations}

In contrast to the electrical quantum dot case which was considered
in Ref.~\onlinecite{ma08} now the effective potential in
Eqs.~(\ref{twosysrad}) is a continuous function at the dot border.
For this reason the boundary conditions (\ref{dbc}) are
equivalent to
\begin{equation}\label{dbcm}
  b_I(r_0) = b_{II}(r_0), \quad b_{I,r}(r_0) = b_{II,r}(r_0).
\end{equation}
These boundary conditions are identical to those for the previous
Schr\"{o}dinger electron case (\ref{bc}). It enables us to use
the full analogy with the previous case. Taking this analogy into
account we have for the solution in the two regions
\begin{subequations}\label{solinout}
\begin{eqnarray}
&&  b_I(r) = Af(r) = Ar^{|m+1|}e^{-r^2/4}M(a_0|c_0|r^2/2), \phantom{mm} \\
&&  b_{II}(r) = BJ_{\nu}(kr) + CY_{\nu}(kr).
\end{eqnarray}
\end{subequations}
where $k = |E|$, $a_0 = (|m+1|+m + 1 - E^2)/2$, $\nu = m+1+r_0^2/2$,
and $c_0 = |m+1|+1$. The expressions for the other wave function
component $a(r)$ follow directly from Eq.~(\ref{twosysrad1}):
\begin{subequations}\label{solinouta}
\begin{eqnarray}
&&  a_I(r) = \frac{A}{E}r^{|m+1|}e^{-r^2/4} \nonumber \\
&& \phantom{m}\times\left\{\frac{d}{dr} +
\frac{|m+1|+m+1}{r}\right\}
  M(a_0|c_0|r^2/2), \phantom{mm} \\
&&  a_{II}(r) = BJ_{\nu-1}(kr) + CY_{\nu-1}(kr).
\end{eqnarray}
\end{subequations}

\begin{figure}[ht]
\begin{center}
\includegraphics[width=7cm]{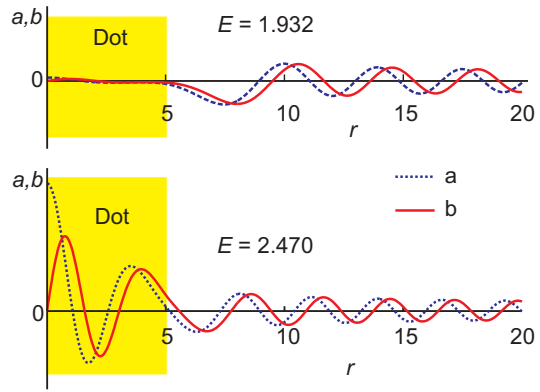}
\caption{(Color online) The wave function components: $a$ -- dashed
blue curve, and $b$ -- red solid curve, for $m=0$, dot radius
$r_0=5$ and two energy values: $E = 1.932$ -- upper plot, $E =
2.470$ -- lower plot.} \label{fig7}
\end{center}
\end{figure}

The wave function components obtained in the above way are
illustrated in Fig.~\ref{fig7} for two different values of the
energy.
We see the same tendency. When the energy is close to the Landau
level energy of the Dirac electron in a homogeneous magnetic field,
\begin{equation}\label{dll}
  E_{n,m} = \pm\sqrt{2n + |m+1| + m + 1},
\end{equation}
(see the lower plot of Fig.~\ref{fig7} where the energy is close to
the Landau level with $m=0$, $n=2$) we see a clear accumulation of
the wave function components in the dot, what indicates a
quasi-bound state.

\subsection{Local density of states}

Developing further the analogy of Eq.~(\ref{equa2}) with the
considered previously case we calculated the local density of states
using Eqs.~(\ref{constpq}), (\ref{constabc}) and (\ref{lds}).
Because we have now a wave function with two components,
Eq.~(\ref{lds}) is modified into
\begin{equation}\label{ldsd}
  \m{I}(E) = 2\pi\int_0^{\infty}rdr F(r)\left\{|a(r)|^2 + |b(r)|^2\right\}.
\end{equation}
Now the normalization factor is $N=k/4R$ (due to the two wave
function components), and the factor responsible for the change of
the summation over discrete eigenvalues into an integral over the
electron energy is $R/\pi$. Thus, the local density of states in the
case of a Dirac electron becomes
\begin{equation}\label{dirlds}
  \rho(E) = \frac{|E|}{2}\int_0^{\infty}rdr f(r)
  \left\{|a(r)|^2 + |b(r)|^2\right\}.
\end{equation}

In the case of free Dirac electrons (when there is no magnetic dot)
the wave function components read
\begin{equation}\label{dirfree}
  a_{\mathrm{free}} = J_{m}(kr), \quad b_{\mathrm{free}} = J_{m+1}(kr),
\end{equation}
what leads to the following expression of the local density of
states for a free electron:
\begin{equation}\label{dflds}
  \rho_{\mathrm{free}}(E) = \frac{|E|r_0^2}{4}e^{-E^2/2b}
  \left\{I_m(E^2/2b) + I_{m+1}(E^2/2b)\right\}.
\end{equation}

\subsection{Numerical results}

\begin{figure}[ht]
\begin{center}
\includegraphics[width=7cm]{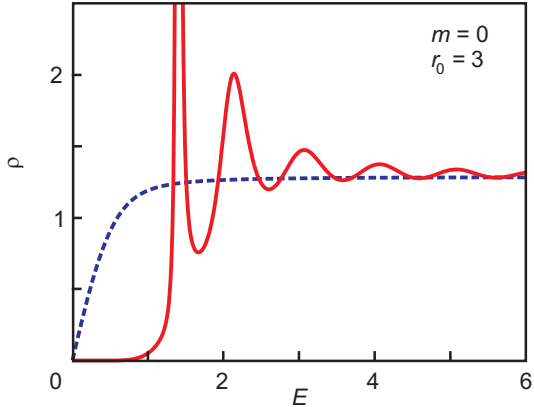}
\caption{(Color online) The local density of states for a Dirac
electron in the magnetic dot for $m=0$ and $r_0 = 3$ shown by red
solid curve. The dashed blue curve is the free electron density of
states.} \label{fig8}
\end{center}
\end{figure}

The typical local density of states calculated for $m = 0$ and $r_0
= 3$ is shown in Fig.~\ref{fig8} for positive energies.
Two differences with respect to standard electrons can clearly be
noticed. First, in the case of the Dirac electron  the spectrum is
symmetric with respect to energy inversion ($E\to -E$) due to the
equivalence of electrons and holes. Thus the plot in Fig.~\ref{fig8}
has to be supplemented by the same curves for negative energies.
Second, comparing the density of states for Dirac electron with the
same curve for the Schr\"{o}dinger one (see Fig.~\ref{fig3}) we see
that there are more peaks. This can be explained by the more dense Landau
level spectrum in the case of Dirac electrons (\ref{dll})
for the large quantum number values
as compared with these for the previous case (\ref{ll}).

\begin{figure}[ht]
\begin{center}
\includegraphics[width=8cm]{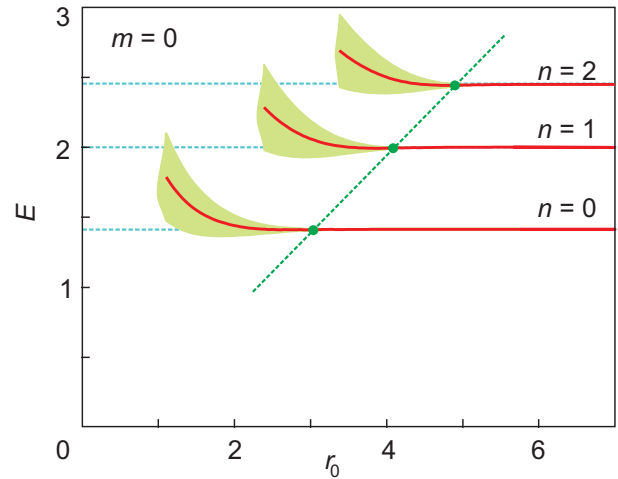}
\caption{(Color online) Quasi-bound states with orbital momentum
$m=0$ for the Dirac electron in the magnetic dot. The energy of
these states are given by red solid curves and its width (i.~e.~the
inverse of the life-time) by the shadowed regions. The Landau levels
are indicated by blue dashed lines.} \label{fig9}
\end{center}
\end{figure}

\begin{figure}[ht]
\begin{center}
\includegraphics[width=8cm]{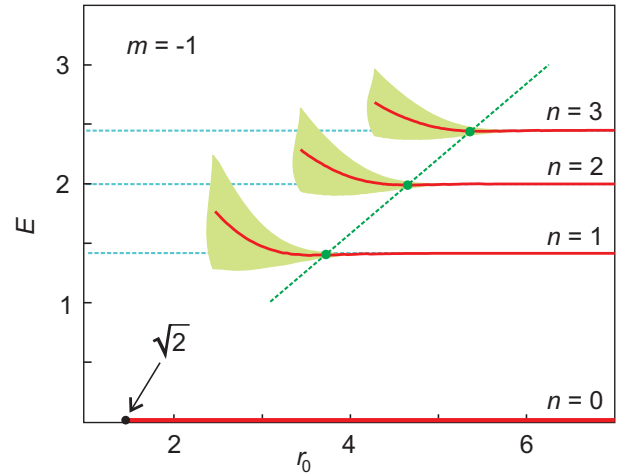}
\caption{(Color online) The same as Fig.~\ref{fig4} but now for
$m=-1$.} \label{fig10}
\end{center}
\end{figure}

As before we fit the peaks by Lorentz type curves what leads to the
broadened levels displayed in Figs.~\ref{fig9} and \ref{fig10}.

We see that the levels with $m=-1$ start at lowest energies what is
just the consequence of chosen definition of radial wave function
components (\ref{wfdir}).

The green dotted curves divide the region of broadened quasi-bound
states from the region where the states have a very small
broadening. These curves were obtained in the same way as it was
done in section \ref{compentex} for the case of Schr\"{o}dinger
electron, namely, applying the complex energy eigenvalue technique.
In the case of the Dirac electron it leads to the following
imaginary energy part:
\begin{equation}\label{denim}
  \Delta = \frac{H_{\nu-1}b - H_{\nu}a}{H_{nu}a_r + H_{\nu,r}a
  -H_{\nu-1}b_r-H_{\nu-1,r}b}.
\end{equation}
The above mentioned dotted green line corresponds to
$\delta=3\cdot10^{-3}$. Note we chose it three times smaller than in
Figs.~\ref{fig4}-\ref{fig6} what causes us to conclude that between
the local density of states technique and complex energy eigenvalue
method one can expect only qualitative agreement.

There is one more interesting point --- the zero energy state which
is shown in Fig.~\ref{fig10} by thick red line along the $x$-axis.
Its behavior differs essentially from all other states. That is why
it needs some special attention which is presented in the next
subsection.

\subsection{Zero energy state}

Now we check whether the Dirac electron has a zero energy state in
the magnetic dot. In this case instead of Eqs.~(\ref{twosysrad}) we
have to solve the following two equations for the radial components
of the electron wave function:
\begin{subequations}\label{zerosys}
\begin{eqnarray}
\label{zerosys1}
  \left\{\frac{d}{dr} + A(r) + \frac{m+1}{r}\right\}b &=& 0, \\
\label{zerosys2}
  \left\{\frac{d}{dr} - A(r) - \frac{m}{r}\right\}a &=& 0.
\end{eqnarray}
\end{subequations}
These are uncoupled differential equations of the first order, and
their solution can be found by a straightforward integration. The
solution has the following asymptotic behavior:
\begin{equation}\label{zeroa}
\begin{split}
  \ln a(r) &= \int dr \left\{A(r) + \frac{m}{r}\right\} \\
  &\sim \left\{\begin{array}{ll}
    m\ln r + r^2/4, & r \to 0; \\ (m +r_0^2/2)\ln r, & r \to \infty,
  \end{array}\right.
\end{split}
\end{equation}
and
\begin{equation}\label{zerob}
\begin{split}
  \ln b(r) &= -\int dr \left\{A(r) + \frac{m+1}{r}\right\} \\
  &\sim \left\{\begin{array}{ll}
    -(m+1)\ln r - r^2/4, & r \to 0; \\ -(m+1+r_0^2/2)\ln r, & r \to \infty,
  \end{array}\right.
\end{split}
\end{equation}
or
\begin{equation}\label{zeroaex}
\begin{split}
  a(r) \sim \times \left\{\begin{array}{ll}
    r^m \exp(r^2/4), & r \to 0; \\ r^{m+r_0^2/2}, & r \to \infty,
  \end{array}\right.
\end{split}
\end{equation}
and
\begin{equation}\label{zerobex}
\begin{split}
  b(r) \sim \left\{\begin{array}{ll}
    r^{-m-1}\exp(-r^2/4), & r \to 0; \\ r^{-(m+1+r_0^2/2)}, & r \to \infty.
  \end{array}\right.
\end{split}
\end{equation}

In order to have the wave function with finite norm two boundary
conditions have to be satisfied. First, the function should behave
like $r^{\alpha}$ ($\alpha \geqslant 0$) when $r \to 0$, and second,
it should behave like $r^{-\alpha}$ ($\alpha \leqslant -1$) when $r
\to\infty$.

For the $a$ component the above conditions reduce to the
requirements $m \geqslant 0$ and $m+r_0^2/2 \leqslant -1$, which can
not be satisfied simultaneously. Consequently, we have to assume
that $a=0$.

In the case of component $b$ the conditions read
\begin{equation}\label{bcond}
  -r_0^2/2 \leqslant m \leqslant -1,
\end{equation}
from which it follows that if $r_0^2/2 \ge 1$ there are always some
negative $m$ values for which a zero energy state exists. When the
radius of the dot decreases this interval becomes smaller, and the
zero energy states vanish one by one. Finally, at $r_0^2/2 < 1$ all
of them disappear.

Such essential difference between the bound zero energy level and
all other quasi-bound levels is caused by the fact, that the wave
function of the state with zero energy is real. Consequently, the
electron in this state has no velocity, and as a result there is no
tunnelling of this electron outside the dot. Unfortunately, the
absence of any non zero electron velocity makes it impossible to
reveal this state in transport measurements, but maybe it can reveal
itself through the statistic properties of the magnetic dot.

\section{Conclusions}

We considered the eigenvalue problem of a model quantum magnetic dot
where the homogeneous magnetic field perpendicular to the 2D
electron motion plane is created only in a finite region
--- in a circle of radius $r_0$. We showed that such a magnetic
field fails to confine electrons both for the standard parabolic
dispersion law or the ultra relativistic linear dispersion law for
Dirac electrons in graphene. Although in such a magnetic dot no
confined states are found, quasi-bound sates with a finite lifetime
are present.

An analysis of the quasi-bound states for the Schr\"{o}dinger and
Dirac electron was performed by means of the local density of
states, and the position and width of the resonance peaks (the
analogs of the quasi-stationary states), on the dot radius (or the
strength of the magnetic field) were calculated.

The broadening of these peaks (the inverse lifetime of the
quasi-bound state) is mainly caused by the touching of the quantum
dot border by the electron wave function tail. Due to the
exponential character of this tail there exist a rather sharp border
between broadened quasi-bound states and those which can be
considered as nearly bound ones. This border was found by applying
the complex energy eigenvalue method which is shown to be in
qualitative agreement with the results obtained from the local
density of states technique.

It is shown that the difference of the quasi-bound states in the
magnetic dot between the Schr\"{o}dinger and Dirac electrons is only
in the energies of these states which is a consequence of the
different energies of the corresponding Landau levels.

There is a single exception: in the case of Dirac electrons there
exists a zero energy bound state for negative values of the angular
momentum (the momentum which is opposite to the direction of the
classical electron rotation along the Larmor circle). When the dot
radius $r_0$ (or the magnetic field strength) decreases the
degeneracy of this zero energy level decreases skippingly while all
other quasi-bound states disappear smoothly via their broadening.

\onecolumngrid

\begin{acknowledgments}
This work was supported by the Flemish Science Foundation (FW0-Vl),
CNPq (Brazil science funding agency) and the Belgian Science Policy
(IAP).
\end{acknowledgments}

\end{document}